# Solvation of the Morpholinium Cation in Acetonitrile. Effect of an Anion

Vitaly V. Chaban[1] and Nadezhda A. Andreeva[2]

[1] Instituto de Ciência e Tecnologia, Universidade Federal de São Paulo, 12231-280, São José dos Campos, SP, Brazil.

[2] Unemployed.

**Abstract**. Ionic liquids constitute a fast growing class of compounds finding multiple applications in science and technology. Morpholinium-based ionic liquids (MBILs) and their mixtures with polar molecular co-solvents are interesting as sustainable electrolyte systems for electrochemistry. We investigate local structures of protic and propton-free morpholinium cations in acetonitrile (ACN) using semi-empirical molecular dynamics (MD) simulations. An impact of an anion (acetate) on the cation solvation regularities is discussed. Unlike oxygen, nitrogen of the morpholine ring is a strong electrophilic binding center. This site is responsible for the interactions of the cation with the solvent and with the anion. In protic MBILs, the role of nitrogen is delegated to the proton, which is linked to nitrogen. The acetate anion weakens solvation of the cation due to occupation of space near nitrogen or proton. The analysis reveals a favorable solvation of MBILs in ACN, which is a prerequisite for a new high-performance electrolyte system. The reported structural data are validated through a point-to-point comparison with the MP2 post-Hartree-Fock theory.

**Introduction**

Ionic liquids (ILs) constitute an established class of compounds with unique physical chemical properties.[1-13] These properties include negligible vapor pressure, high electrical conductivity, electrochemical stability, non-flammability, tunability and unusually large temperature range of the liquid state. ILs of current research interest[1] can be constructed out of cations, representing imidazolium, pyridinium, pyrrolidinium, ammonium, phosphonium, cholinium, and morpholinium families. These cations can be coupled with the universe of anions, e.g. chloride, bromide, acetate, tetrafluoroborate, hexafluorophosphate, alkylsulphate, nitrate, thiocyanate, bis(triluorosulfonyl)imide, etc. Certain ILs exhibit a low toxicity and can partially or fully substitute conventional molecular solvents in multiple applications. Molecular and ionic liquids are also able to efficiently work together to allow a very fine tuning of physical chemical properties.[1] In this context, emergence of new ILs is important and exciting.

Morpholinium-based ILs (MBILs) are relatively new. Their investigation – thus far – has been far less intensive,[14-30] as compared to the imidazolium- and pyridinium-based ILs. MBILs are likely of a great practical importance. They were shown to perform well in the electrolyte systems for lithium batteries, supercapacitors and solar cells. Simple synthesis procedures of MBILs,[29] cheap precursors of the morpholinium cation and, arguably, a high lithium ion mobility fostered by the oxygen site are favorable factors. MBILs exhibit a large electrochemical window and a relatively large ionic conductivity.[26] Both morpholine and morpholinium salts are miscible with water. Unfortunately, high shear viscosities of pure MBILs complicates a straightforward application. According to Galinski and Stepniak, mixtures of MBILs and propylene carbonate are prospective electrolytes for electrochemical double-layer supercapacitors.[27]

Lava and coworkers[28] investigated crystals, which morpholinium-based cations form with sulfosuccinate anions. Long alkyl chains (8-18 carbon atoms) were found to favor hexagonal columnar phases at room conditions. The morpholinium-based cations structurally resemble the piperidinium-based cations by mass, shape and substituents. The positive charge center coincides with the nitrogen atom, although electron deficiency is significantly delocalized over the ring. Nevertheless, the presence of the oxygen atom puts morpholinium aside. Since electronegativity of oxygen is higher than electronegativity of nitrogen, an electron-rich interaction site may be expected within the cation. This can favor specific interactions of morpholinium with other electrochemically relevant cations, including the lithium ion. It is unclear to which extent these interactions can be implemented, since the other part of the morpholinium cation is positively charged and should naturally repel other cations.

As known from Lemordant and coworkers,[26] MBILs are viscous. High viscosity makes liquids difficult to work with and pre-determines a small self-diffusion of charge carriers. Addition of a polar molecular co-solvent is able to screen electrostatic interactions between the cation and the anion. Although this alteration of composition decreases a concentration of charge carriers, it exponentially increases self-diffusion.[31] Since a linear concentration decrease is accompanied by an exponential viscosity decrease, an overall effect on the ionic conductivity is favorable. Normally, addition of a small portion of a polar molecular co-solvent leads to certain increase of conductivity. The conductivity reaches maximum around 10-15 mol% IL in the mixture and decreases drastically towards zero upon further dilution.[31,32] Knowledge about the position and the value of the conductivity maximum appears of a great practical importance[32,33] for electrochemical applications. Understanding energetics, structure and dynamics of the IL-co-solvent mixtures fosters development of robust and efficient electrolyte solutions for energy storage, conversion and production applications.[4-6,9,10]

Here, we report numerical simulations of the protic and proton-free morpholinium-based cations – N-ethyl-N-methylmorpholinium (EMM) and N-methylmorpholinium (MM) – were considered in relation to their solvation in acetonitrile (ACN). Furthermore, an impact of an anion (acetate, AC) on the solvation regularities of the above mentioned cations was investigated. We chose ACN as a co-solvent due to its wide applications in electrochemistry, a high self-diffusion coefficient and a high dipole moment. Acetate was chosen deliberately, since it is known to form stable cation-anion pairs with many cations. Solvation of the morpholinium containing ion pair is our primary research interest.

**Methodology**

Structural data reported in the present work were derived from the ion-molecular trajectories following thorough equilibration. The 200 ps long ion-molecular trajectories at 300 K were obtained using the PM7-MD simulations.[34-39] Constant temperature was maintained using the Andersen coupling scheme.[40] Integration time-step was set to 0.0005 ps, irrespective of the system identity. Frames for further statistical mechanical analysis were saved every 0.02 ps. At every time-step, forces on each atom were obtained using the PM7 method with a wave function convergence threshold of $10^{-6}$ Ha. The systems were represented as clusters in vacuum, i.e. neglecting periodic boundary conditions.

PM7 is the latest semi-empirical method,[34] in which the Hartree-Fock secular equation, |H–ES| = 0, is replaced with a simpler equation, |H–E|=0. This approximation is known as a neglect of diatomic differential overlap (NDDO). Only the valence electrons are treated quantum mechanically, while the lower-energy electrons are accounted for via the core potentials. Thanks to a careful parametrization of NDDO using multiple experimental data, PM7 implicitly includes electron-correlation effects and reproduces other quantum phenomena. Longer-range weak interactions (van der Waals forces) are included by means of a separate empirically-based re-optimized function.

PM7-MD is an ideal tool for simulation of many-component systems, in which specific interactions – such as hydrogen bonding, π-stacking, electronic polarization, etc – are anticipated. A significant size of the many-component systems (expressed either in atoms or in electrons) prohibits an application of density functional theory or higher-level quantum-chemical methods to simulate molecular dynamics (MD). In turn, classical MD simulations, normally utilizing simple and computationally efficient pairwise interactions, miss many important effects, unless the parameters were deliberately developed for those systems. Unfortunately, cost of parametrization increases exponentially with the number of atom types in the system.

PM7-MD was successfully applied to nanostructures,[36,41] gas capture,[42] ion-molecular systems,[39] competitive solvation,[35,37] achieving quantitative agreement with the available experimental data, where applicable. Accuracy of PM7-MD applying particularly to the present work was validated through a parallel description of a simple ion-molecular system by the M11 hybrid density functional theory (HDFT),[43] the Møller-Plesset second-order perturbation theory (MP2),[44] and PM7.[34] The MP2 and M11 calculations used the atom-centered, split-valence, triple-zeta, polarized, Pople 6-311+G* basis set. Inclusion of diffuse functions ('+') to the basis set is recommended to adequately represent anions.

**Results and Discussion**

We constructed four major MD systems, not counting a few MD systems used for benchmarking and accuracy validations: (1) $EMM^+$ immersed into 16 ACN molecules; (2) $EMM^+$ $AC^-$ ion pair immersed into 16 ACN molecules; (3) $MM^+$ immersed into 16 ACN molecules; (4) $MM^+$ $AC^-$ ion pair immersed into 16 ACN molecules. The $MM^+$ cation constitutes a particular interest, since it is protic. We anticipate that hydrogen linked to nitrogen represents an additional coordination site both for the anion and the solvent molecules. 16 ACN were chosen to reproduce a compete first solvation shell (FSS) and, partially, a second solvation shell. Unlike in the case of smaller cations, the second shells in ILs are either poorly defined or absent due to a smaller (delocalized) charge density. Although our models do not meet all the criteria of liquid phase, the cations are located far from the surface. Therefore, their properties can be directly compared to those in the periodic systems above the thermodynamic limit.

According to Figure 1, which visualizes immediate equilibrated ion-molecular configurations, the MM and EMM cations are located at the center of the system, whereas the acetate anion is solvated to a lesser extent. This is in line with the classical chemical knowledge, as anions are generally solvated weaker than cations.

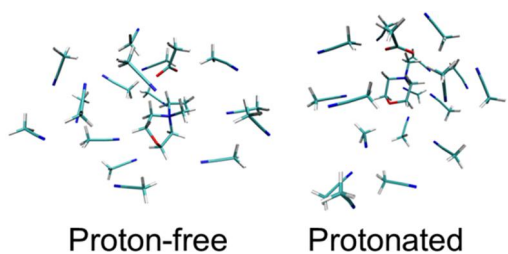

Figure 1. The equilibrated ion-molecular configurations of the simulated systems containing the proton-free (left) and protic morpholinium-based cations. Carbon atoms are cyan, hydrogen atoms are white, oxygen atoms are red, nitrogen atoms are blue. VMD[45] was used to render the output of the in-home PM7MD code.

The equilibration of each of four primary systems was thoroughly controlled based on the evolution of multiple thermodynamic quantities. Three of them, potential energy, kinetic energy and temperature, are provided in Figure 2 vs. simulation time. The systems equilibrate very quickly, thanks to their relatively small size and the absence of a long-range structure. Furthermore, ACN molecules are mobile exhibiting a high self-diffusivity, ca. $4.3 \times 10^{-9}$ m$^2$ s$^{-1}$, at room conditions. The simulations were started from the cold configurations, which were promptly heated via coupling to the thermostat, as described in the methodology. Heating up to 300 K takes ca. 2.0 ps, after which temperature fluctuates around 300 K. The observed fluctuation of temperature is commensurate with the size of the system (number of degrees of freedom).

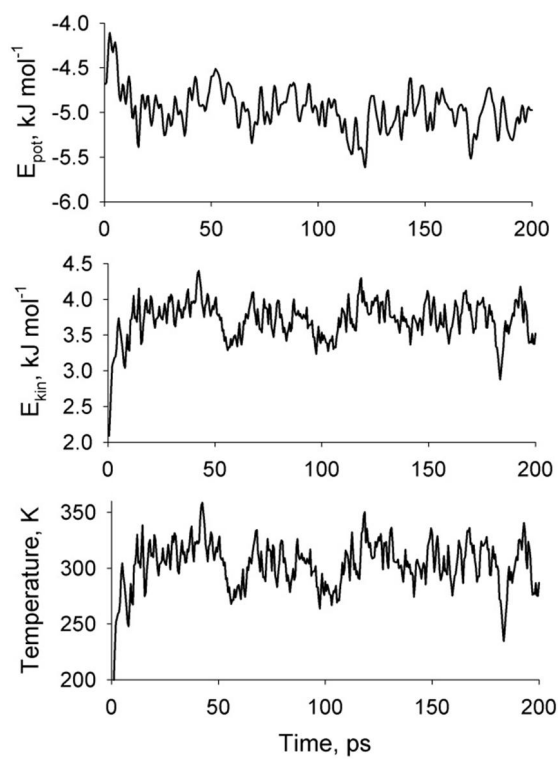

Figure 2. Evolution of potential energy, kinetic energy and translational temperature in the course of the simulations. The data correspond to the system containing EMM, AC and 16 ACN. Note that the depicted energies were normalized per mole of atoms.

Figures 3-7 provide a comprehensive description of the atom-atomic structure and preferential coordination of MM and EMM in ACN, in the presence of the acetate anion and in the free state (the infinite dilution approximation). Simultaneous analysis of these plots is, in principle, enough to retrieve exhaustive information regarding the existence of these mixtures. The oxygen and nitrogen anticipated solvent and anion coordination sites are compared in Figure 3 (for the protic cation, MM) and in Figure 4 (for the proton-free cation, EMM). Coordinating potential of the oxygen atom in the morpholine ring is weak. This feature does not essentially change upon formation of the ion pair, although shapes and positions of the peaks somewhat change. It is clear that oxygen is a negatively charged (electron-rich) interaction site, since it exhibits a stronger peak with the methyl group of ACN. The methyl group of ACN holds certain positive charge, whereas nitrogen is negatively charged. The presence of a positively charged center in the ring of the cation is a unique feature making the morpholinium-based ILs

particularly unusual. Although it can be concluded from ab initio description in vacuum that the oxygen atom of the morpholine ring is electron-rich, this fact does not implicitly mean that the atom retains this feature in all solvents and paired with all anions. Maintaining oxygen as a partially negative atom is believed to be important for the applications in lithium-batteries. The lithium cation is anticipated to bind oxygen to certain extent, although this feature remains to be proven.

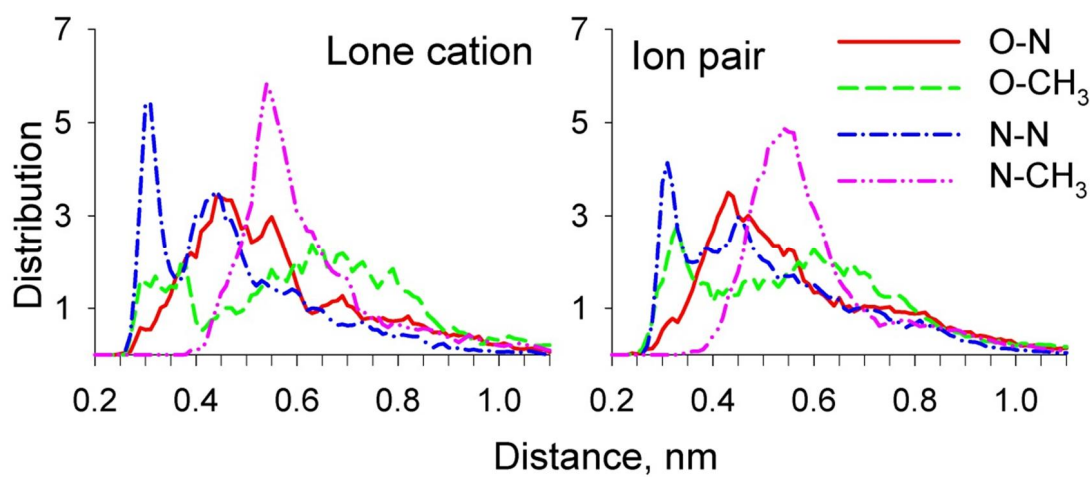

Figure 3. Distance distribution functions computed for the selected pairs of interaction centers: O, N (MM) and N, CH$_3$ (ACN).

A strong interaction is observed between the nitrogen atom of MM and the nitrogen atom of ACN (Figure 3). This correlation weakens (5.5 vs. 4.5 units) upon acetate addition. A strong structural correlation between the cation and the solvent is favorable to accelerate diffusivity of the system. One anticipates that the mobile ACN molecules greatly immobilize cations, following the previously exemplified mechanism,[31] since an ionic self-diffusion largely depends on a particular microscopic environment. Furthermore, the nitrogen-nitrogen binding is responsible for a good mutual miscibility of the MM-based ILs and ACN.

Presence or absence of proton in the morpholinium cation plays an important role. Figure 4 is quite different as compared to Figure 3, although the only difference between MM and EMM is the proton linked to nitrogen. For instance, the absence of proton weakens the above discussed nitrogen-nitrogen coordination. Interestingly, the oxygen-solvent binding becomes somewhat stronger.

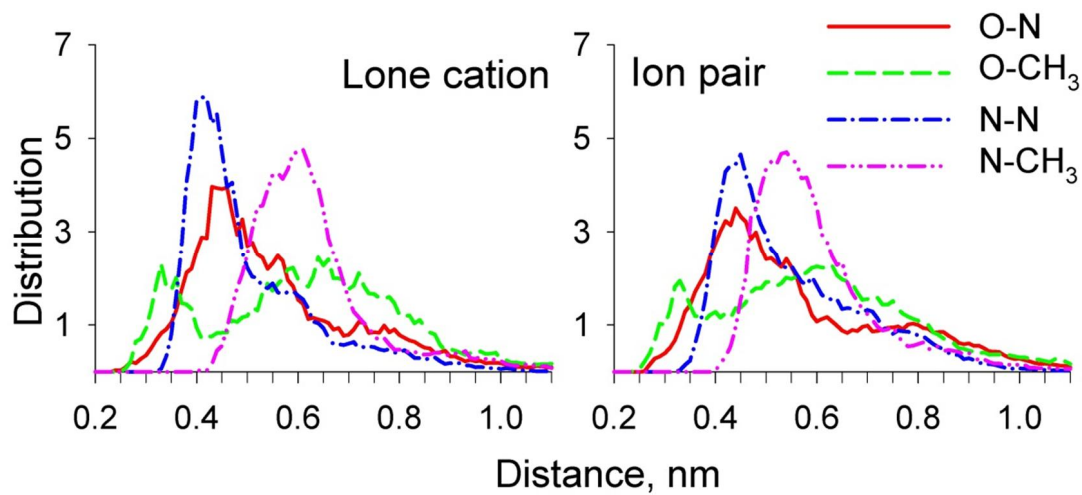

Figure 4. Distance distribution functions computed for the selected pairs of interaction centers: O, N (EMM) and N, CH$_3$ (ACN).

The methyl group of EMM does not constitute an attractive binding center for ACN (Figure 5). Both distributions for a positively (CH$_3$) and negatively (N) charged sites of ACN with the methyl group of EMM are smashed. In turn, the proton of MM binds ACN strongly (peak height of 6). This strong binding is also supported by the H-Me(ACN) peak at 0.46 nm. The acetate anion, once present near MM, ruins this noteworthy peak. The cation-anion structure regularities will be in detail discussed below.

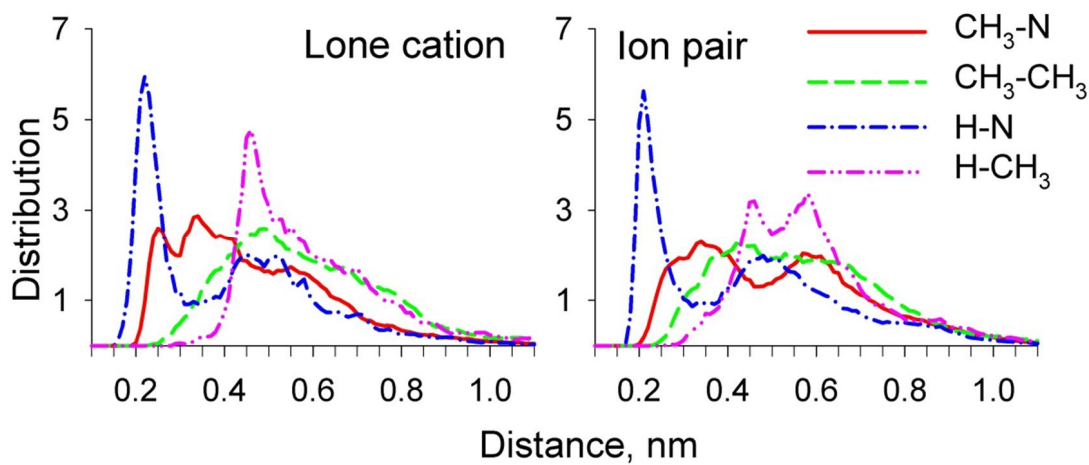

Figure 5. Distance distribution functions computed for the selected pairs of interaction centers: $CH_3$ (EMM), H (MM) and N, $CH_3$ (ACN).

Figure 6 investigates carbon atoms of the morpholine ring. We do not separate the four carbon atoms of the ring, although the 2 and 6 positions, strictly speaking, are not symmetrically equivalent to the 3 and 5 positions. All those carbon atoms are slightly electron deficient and favor a relatively modest binding with the nitrogen atom of ACN. Correlations with the methyl group of the solvent are not pronounced. As it was also observed in the previous cases, addition of an anion prevents a solvent molecule coordination, although not drastically. This constitutes indirect marker that the acetate anion tends to be located in the vicinity of MM and EMM.

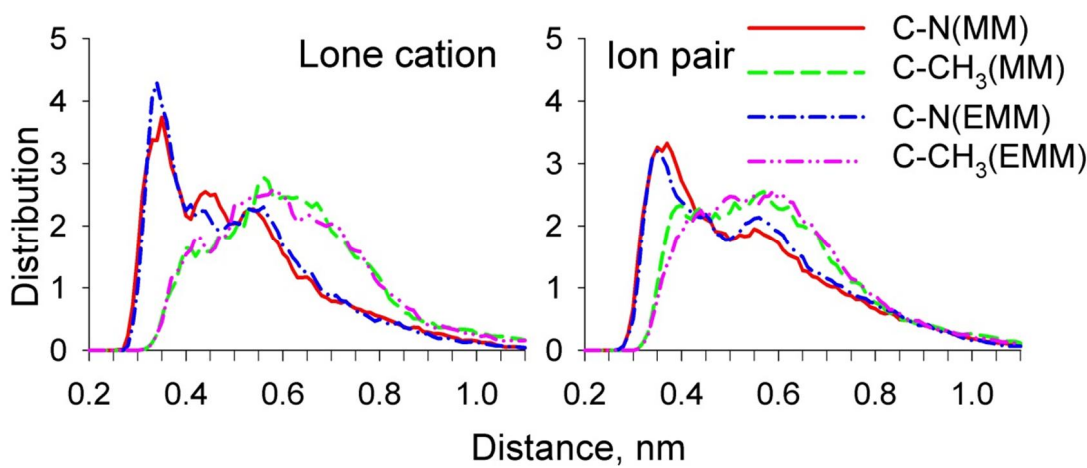

Figure 6. Distance distribution functions computed for the selected pairs of interaction centers: C (MM and EMM) and N, $CH_3$ (ACN).

Indeed, the cation-anion correlations are stronger (Figure 7) than those between the cation and the solvent. As may be expected, the carboxyl group provides a driving force for binding EMM. An electrostatic attraction between nitrogen of EMM and oxygen of AC is primarily responsible for the presence of AC in the FSS of EMM. This supposition perfectly agrees with the attenuating correlation peaks in the case of the anion addition. It deserves consideration that the oxygen-oxygen peak is unexpectedly well-shaped, whereas the methyl-oxygen pair does not exhibit any comparable correlation.

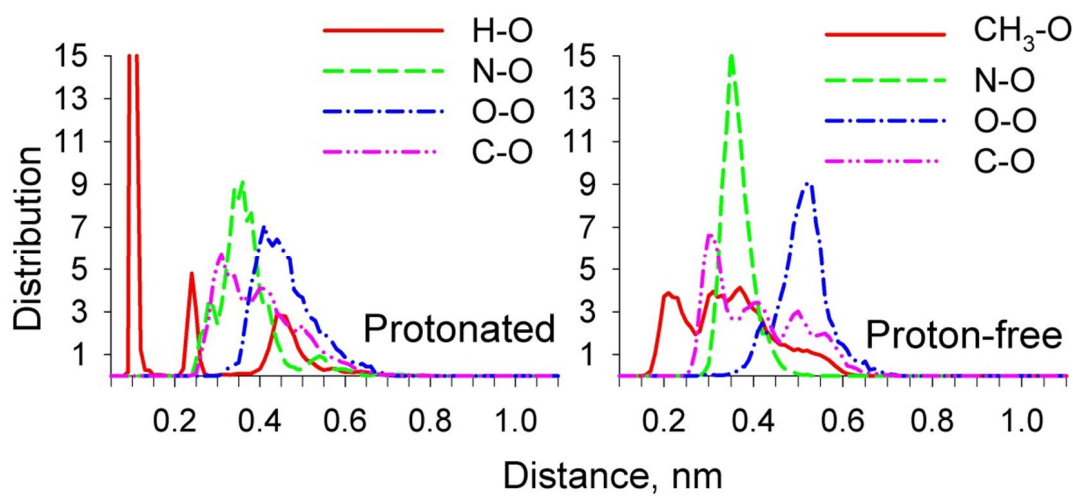

Figure 7. Distance distribution functions representing interactions between the MM and EMM cations (H, N, C, $CH_3$) and the acetate anion (O).

According to Figure 7, the proton of MM and oxygen of AC form a true covalent bond, ca. 1.0 Å long, with a peak height of 44 units. This bond is not stable, however. The second O-H peak is at 0.24 nm (5 units). This one corresponds to a distance between the proton and another oxygen atom of AC. The third, wider, peak at 0.44 nm (3 units) reflects a non-bonded state. That is, the proton migrates between nitrogen and oxygen and does not belong to any of them during an entire simulation time. Once the proton is cleaved, the AC anion transforms into acetic acid and EM transforms into a molecule. Nevertheless, such a molecule is of low stability and possesses a doublet electronic configuration in the ground state. As a result of thermal motion, the proton fluctuates in between MM and AC. To confirm the discussed observations, the M11

HDFT calculation was performed (Figure 8). The calculation fully confirms our interpretation, including the implemented lengths of the covalent and hydrogen bonds, with a correction for thermal expansion and thermal motion. Although internal energy in vacuum suggests location of hydrogen within the carboxyl group (indeed, oxygen is more electronegative), other components of the Gibbs free energy, likely, favor electrolytic dissociation of AC. Furthermore, an effect of solvent must be considered, which we did not include in the simplified HDFT calculations (Figure 8).

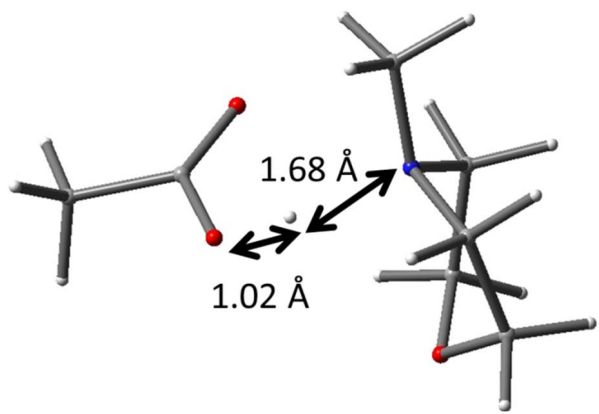

Figure 8. The M11 hybrid density functional theory prediction of the minimum energy configuration. The simulated system consists of the 1-methylmorpholinium cation and the acetate anion. AC pulls proton towards itself engendering a covalent bond, whereas nitrogen maintains a strong hydrogen bond, 1.68 Å.

One may question accuracy of the discussed PM7-MD simulations, as compared to more traditional electronic-structure simulation methods, such as density functional theory. Table 1 provides a detailed comparison between the results obtained for the same nuclear-electronic system, but using three different methods: PM7-MD, M11 hybrid density functional theory, and Møller-Plesset second-order (MP2) perturbation theory, the post-Hartree-Fock electron-correlation method. The geometry of the system was independently optimized using the same convergence criterion and three different electronic-structure methods. After that, a point-to-point comparison summarized in Table 1 was performed. Out of the presently applied electronic-structure methods, MP2 is the highest level of theory and the most computationally expensive method. It is to be used as a reference, should significant discrepancies of PM7 and M11 occur.

Table 1. Comparison of PM7, M11 HDFT, and MP2, as applied to the [EMM+AC+ACN] neutral system. The most structurally important distances and angles were selected for comparison. In the cases where a few equivalent distances were available for the same pair of atoms, the shortest one was used by every method

| Designation of interaction | Distances, Å / angles, degrees | | |
|---|---|---|---|
| | PM7 | M11 HDFT | MP2 post-Hartree-Fock |
| Intra-molecular | | | |
| C-N(EMM) | 1.51 | 1.51 | 1.52 |
| C-C(EMM) | 1.54 | 1.54 | 1.52 |
| C-O(EMM) | 1.41 | 1.42 | 1.42 |
| N-C(EMM) | 1.52 | 1.49 | 1.52 |
| C-H(EMM) | 1.10 | 1.10 | 1.09 |
| C-O-C(EMM) | 113 | 111 | 109 |
| C-N-C(EMM) | 109 | 107 | 109 |
| C-N(ACN) | 1.16 | 1.15 | 1.18 |
| C-C(ACN) | 1.43 | 1.46 | 1.46 |
| C-O(AC) | 1.25 | 1.26 | 1.26 |
| C-C(AC) | 1.52 | 1.53 | 1.53 |
| O-C-O(AC) | 124 | 125 | 124 |
| Inter-molecular | | | |
| N(EMM)-O(AC) | 3.26 | 3.36 | 3.28 |
| O(EMM)-O(AC) | 4.25 | 3.96 | 4.18 |
| C(CH$_2$,EMM)-O(AC) | 2.94 | 3.14 | 2.29 |
| N(EMM)-N(ACN) | 4.28 | 4.57 | 4.31 |
| O(EMM)-N(ACN) | 5.42 | 4.54 | 4.49 |
| O(AC)-C(CH$_3$,ACN) | 2.96 | 3.10 | 3.13 |
| N(EMM)-O(EMM)-O(AC) | 43 | 43 | 43 |
| N(ACN)-N(EMM)-O(AC) | 57 | 64 | 57 |

All three methods exhibit a decent performance being applied to intra-molecular geometry parameters. Note that intra-molecular parameters were analyzed in the optimized complex geometry, rather than in the isolated states. In five cases out of twelve, predictions of M11 are closer to those of MP2. In four cases, PM7 provides a higher accuracy than M11, although M11 is much more computationally expensive. Note, however, that the discussed differences are marginal and never exceed 0.03 Å and 2 degrees. PM7 also appears quite successful in the case of intermolecular interactions. Since these interactions are not valence, a correct description of

Coulombic interactions, a weak dispersion attraction, adequate atomic radii and other related properties are of critical importance. According to Table 1, an empirical parametrization appears quite beneficial, because the results of PM7 are in many cases better (closer to MP2) than those of M11. The observed discrepancies among the three different methods are larger, as compared to those for the valence parameters. To summarize, the comparison of PM7 to HDFT does not reveal any systematic faults preventing further application. Consequently, we conclude that the reported structural results are comparable in reliability to the higher-level of theory and can be fully trusted.

**Conclusions**

To recapitulate, the PM7-MD simulations were carried out for the relatively large systems involving the morpholinium-based cation, the acetate anion, and 16 ACN molecules. An electronic-structure level of theory allows to capture all specific interactions in these systems and unravel their local structures.

The role of the morpholine oxygen atom as a binding site is modest, both in relation to the solvent and the anion. While such a behavior with the anion was predictable, it had to be demonstrated for the polar co-solvent molecule. In turn, nitrogen in the proton-free MBIL and hydrogen in the protic MBIL perform strongly. The presence of the acetate anion weakens solvation due to formation of a stable ion pair. An atomistically precise investigation of the lithium cation coordination in the vicinity of the protic and proton-free morpholinium-based cations remains to be performed. Our present results foster development of new electrolytes employing ion-molecular systems. Knowledge about preferential coordination centers is necessary to suggest new anions, co-solvents and further functionalize morpholinium cations.

## Acknowledgments

V.V.C. is CAPES fellow.

## Author Information

E-mails for correspondence: vvchaban@gmail.com (V.V.C.); nadejda-spb@list.ru (N.A.A.)

TOC Graphic

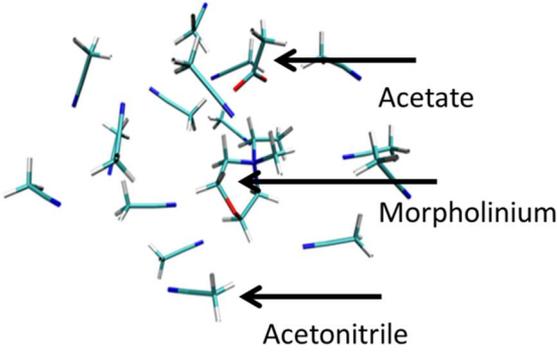